\documentclass[12pt]{article}
\usepackage{a4wide}
\usepackage{latexsym}
\usepackage{amsmath}
\usepackage{amsfonts}
\usepackage{amssymb}
\usepackage{amscd}
\usepackage{cite}
\usepackage{placeins}

\usepackage{pslatex}
\usepackage{graphicx}
\usepackage[latin1,utf8]{inputenc}
\usepackage[T1]{fontenc}

\allowdisplaybreaks

\newcommand{\bq}{\begin{eqnarray}}
\newcommand{\eq}{\end{eqnarray}}
\newcommand{\eps}{\varepsilon}
\newcommand{\qbar}{\bar{q}}
\newcommand{\path}{{\mathcal C}}
\newcommand{\slashoperator}[2]{|_{#2} #1}

\begin{document}

\thispagestyle{empty}

\begin{flushright}
  MITP/20-067
\end{flushright}

\vspace{1.5cm}

\begin{center}
  {\Large\bf Modular transformations of elliptic Feynman integrals\\
  }
  \vspace{1cm}
  {\large Stefan Weinzierl \\
  \vspace{1cm}
      {\small \em PRISMA Cluster of Excellence, Institut f{\"u}r Physik, }\\
      {\small \em Johannes Gutenberg-Universit{\"a}t Mainz,}\\
      {\small \em D - 55099 Mainz, Germany}\\
  } 
\end{center}

\vspace{2cm}

\begin{abstract}\noindent
  {
We investigate the behaviour of elliptic Feynman integrals under modular transformations.
This has a practical motivation:
Through a suitable modular transformation we can achieve that the nome squared is a small quantity, 
leading to fast numerical evaluations.
Contrary to the case of multiple polylogarithms, where it is sufficient to consider just variable
transformations for the numerical evaluations of multiple polylogarithms, 
it is more natural in the elliptic case to consider a combination of a variable transformation 
(i.e. a modular transformation)
together with a redefinition of the master integrals. 
Thus we combine a coordinate transformation on the base manifold with a basis transformation in the fibre.
Only in the combination of the two transformations we stay within the same class of functions.
   }
\end{abstract}

\vspace*{\fill}

\newpage

\section{Introduction}
\label{sect:intro}

Elliptic Feynman integrals in perturbative quantum field theory
and closely related integrals in string theory have received considerable attention 
in recent years \cite{Broadhurst:1993mw,Laporta:2004rb,Bailey:2008ib,MullerStach:2011ru,Adams:2013nia,Bloch:2013tra,Remiddi:2013joa,Adams:2014vja,Adams:2015gva,Adams:2015ydq,Bloch:2016izu,Adams:2017ejb,Bogner:2017vim,Adams:2018yfj,Honemann:2018mrb,Bloch:2014qca,Sogaard:2014jla,Tancredi:2015pta,Primo:2016ebd,Remiddi:2016gno,Adams:2016xah,Bonciani:2016qxi,vonManteuffel:2017hms,Adams:2017tga,Ablinger:2017bjx,Primo:2017ipr,Passarino:2017EPJC,Remiddi:2017har,Bourjaily:2017bsb,Hidding:2017jkk,Broedel:2017kkb,Broedel:2017siw,Broedel:2018iwv,Lee:2017qql,Lee:2018ojn,Adams:2018bsn,Adams:2018kez,Broedel:2018qkq,Bourjaily:2018yfy,Bourjaily:2018aeq,Besier:2018jen,Mastrolia:2018uzb,Ablinger:2018zwz,Frellesvig:2019kgj,Broedel:2019hyg,Blumlein:2019svg,Broedel:2019tlz,Bogner:2019lfa,Kniehl:2019vwr,Broedel:2019kmn,Abreu:2019fgk,Duhr:2019rrs,2019arXiv190811815L,Klemm:2019dbm,Bonisch:2020qmm,Walden:2020odh,Campert:2020yur,Broedel:2014vla,Broedel:2015hia,Broedel:2017jdo,DHoker:2015wxz,Hohenegger:2017kqy,Broedel:2018izr}.
We call a Feynman integral elliptic, if it can be expressed as a linear combination of iterated integrals 
on a covering space of the moduli space ${\mathcal M}_{1,n}$ of
a genus one curve with $n$ marked points with integrands having only simple poles.
``Ordinary'' Feynman integrals, which evaluate to multiple polylogarithms, can
be expressed as a linear combination of iterated integrals 
on a covering space of the moduli space ${\mathcal M}_{0,n}$ of
a genus zero curve with $n$ marked points, again with integrands having only simple poles.

For the numerical evaluation of the iterated integrals on a covering space of the moduli space ${\mathcal M}_{1,n}$
we expand those iterated integrals in a power series in the nome squared $\qbar=\exp(2\pi i \tau)$.
This implies that we first make a choice for coordinates on ${\mathcal M}_{1,n}$, and in particular for the variable
$\tau$, being the ratio of two periods of the elliptic curve.
Any other choice $\tau'$ for this variable is related to the original choice $\tau$ by a modular transformation
from the full modular group $\mathrm{SL}_2({\mathbb Z})$.
By a suitable modular transformation we may therefore achieve that
\bq
 \left| \qbar \right|
 & \le &
 e^{- \pi \sqrt{3}}
 \; \approx \; 0.0043.
\eq
This is a small expansion parameter.

However, if we just consider modular transformations we find that
the iterated integrals on a covering space of the moduli space ${\mathcal M}_{1,n}$
do not transform nicely: We leave the class of functions we started with and generate new integrands
with additional powers of $\ln(\qbar)$.
In particular we may generate negative powers of $\ln(\qbar)$, if integrands of modular weight $0$ and $1$ are present.
This spoils the nice expansion properties.

From a physics point of view this is startling:
We expect that it should not matter which variable we choose as $\tau$ (or equivalently which pair of independent periods we choose
for the elliptic curve).
If we manage to express the Feynman master integrals for one choice of $\tau$ 
nicely as an iterated integral on a covering space of the moduli space ${\mathcal M}_{1,n}$
with integrands from a specific class of integrands, why shouldn't we be able to do so for other choices of $\tau$?

The solution to this riddle is as follows:
We should not only consider a coordinate transformation 
(e.g. going from $\tau$ to $\tau'$ by a modular transformation),
but at the same time also a redefinition of the master integrals.
In fact, we should view a specific choice of master integrals to be tied to a specific choice
of coordinates.
By considering at the same time a coordinate transformation and a redefinition of the master integrals we stay
within the initial class of iterated integrals and do not introduce additional powers of $\ln(\qbar)$.

In this paper we investigate the behaviour of elliptic Feynman integrals under modular transformation in detail.
We explain the need for a redefinition of the master integrals.
This shouldn't come as a surprise. In order to define master integrals of uniform weight in the elliptic case we rescale some
Feynman integrals by a period of the elliptic curve. This involves a choice for this period.
In this way our definition of master integrals of uniform weight is tied to our initial choice of periods (or a choice for the coordinate $\tau$).
If we change our choice of periods (which is equivalent to going from $\tau$ to $\tau'$ by a modular transformation)
we should at the same adapt the definition of the master integrals accordingly.
In the simplest example of the equal mass sunrise integral 
this has already been discussed in ref.~\cite{Honemann:2018mrb} for modular transformations from the congruence subgroup $\Gamma_1(6)$
and in ref.~\cite{Duhr:2019rrs} for modular transformations from the full modular group $\mathrm{SL}_2({\mathbb Z})$.
In the present paper we generalise this observation and allow in particular not only modular forms, but also the coefficients of the Kronecker
function as integrands.

On physical grounds we expect that it should not matter which choice we make for $\tau$.
We usually compute Feynman integrals through the method of differential equations.
Assuming that the system transforms nicely under modular transformations leads to constraints
on the differential equation.
To give an example, consider a system consisting of one master integral $J$ depending on two variables $(z,\tau)$ with differential equation
\bq
 \left( d + A \right) J & = & 0.
\eq
For elliptic Feynman integrals the entry of the $1\times 1$-matrix $A$ is a differential one-form, constructed from modular forms 
and the coefficients $g^{(k)}(z,\tau)$ of the Kronecker function (defined in appendix~\ref{sect:Kronecker}).
We will see that  
\bq
\label{simple_example_1}
 A & = & 
 \eps \left[ g^{(1)}\left(z,\tau\right) - 2 g^{(1)}\left(z,2\tau\right) \right] dz
 + \eps \left[ g^{(2)}\left(z,\tau\right) - 4 g^{(2)}\left(z,2\tau\right) \right] \frac{d\tau}{2\pi i}
\eq
is modular, while the apparent simpler choice
\bq
\label{simple_example_2}
 A & = & 
 \eps g^{(1)}\left(z,\tau\right) dz
 + \eps g^{(2)}\left(z,\tau\right) \frac{d\tau}{2\pi i}
\eq
is not.
Thus requiring modularity of the differential equation restricts the form of the terms which can appear in the matrix $A$.
We call these constraints ``modularity constraints''.
These are additional constraints on top of the integrability constraints 
(both $A$ in eq.~(\ref{simple_example_1}) and eq.~(\ref{simple_example_2}) define a flat connection, 
and hence satisfy the integrability constraint).
This can be used as follows: 
Sometimes Feynman integrals (or their differential equations) are constructed from an ansatz.
If we assume that the elliptic Feynman integrals are modular, we may impose the modularity constraints, leading to fewer
terms in the ansatz.

This paper is organised as follows:
We start with the necessary definitions in section~\ref{sect:definitions}.
In section~\ref{sect:M_1_1} we first discuss elliptic Feynman integrals depending on a single kinematic variable, which 
we may take as $\tau$. In mathematical terms we are considering the moduli space $\mathcal{M}_{1,1}$.
This case is simpler than the general case and serves as a starting point.
In section~\ref{sect:M_1_n} we then discuss the general case of elliptic Feynman integrals depending on $n$ kinematic variables.
In mathematical terms we are now considering a moduli space $\mathcal{M}_{1,n'}$ with $n' \ge n$.
While section~\ref{sect:M_1_1} is essentially restricted to modular forms, we get now in section~\ref{sect:M_1_n}
in addition the coefficients $g^{(k)}(z,\tau)$ of the Kronecker function.
Here, a new complication arises: In the modular transformation of $g^{(k)}(z,\tau)$ terms of lower weight enter with coordinate-dependent coefficients.
In section~\ref{sect:example} we illustrate the general case with a non-trivial example and show the modularity of the 
two-loop sunrise integral with unequal masses.
This is a system with seven master integrals depending on three variables $(z_1,z_2,\tau)$.
In section~\ref{sect:discussion} we return to the general case with a few comments and a discussion.
Finally, our conclusions are given in section~\ref{sect:conclusions}.
The paper is complemented by two appendices:
In appendix~\ref{sect:Eisenstein} we define Eisenstein series for $\Gamma(N)$, 
in appendix~\ref{sect:Kronecker} we define the coefficients $g^{(k)}(z,\tau)$ of the Kronecker function.


\section{Definitions}
\label{sect:definitions}

\subsection{Notation}
\label{sect:notation}

We denote by ${\mathbb H}$ the complex upper half-plane with coordinate $\tau$.
Throughout this paper we use 
\bq
\label{def_basic_variables}
 \qbar & = & \exp\left(2 \pi i \tau \right).
\eq
For a congruence subgroup $\Gamma$ of $\mathrm{SL}_2({\mathbb Z})$ we denote by
${\mathcal M}_k(\Gamma)$ the space of modular forms of weight $k$ for $\Gamma$.
The space of cusp forms is denoted by $\mathcal{S}_k(\Gamma)$,
the Eisenstein subspace is denoted by $\mathcal{E}_k(\Gamma)$.

We denote by $r$ the number of master integrals, by $n$ the number of kinematic variables the master integrals
depend on and by $l$ the number of letters appearing in the differential equation
(i.e. the number of linearly independent differential one-forms).

\subsection{Modular transformation}

Let $\Lambda$ be a two-dimensional lattice in $\mathbb{C}$.
Let $\omega_1$ and $\omega_2$ be two generators of the lattice.
We may assume that $\mathrm{Im}(\omega_2/\omega_1) > 0$, otherwise we simply relabel $\omega_1 \leftrightarrow \omega_2$.
Let $\omega_1'$ and $\omega_2'$ be two other generators, generating the same lattice $\Lambda$.
Then $(\omega_2',\omega_1')$ and $(\omega_2,\omega_1)$ are related by
\bq
\label{def_modular_transformation}
 \left(\begin{array}{c}
   \omega_2' \\
   \omega_1' \\
 \end{array} \right)
 & = &
 \left(\begin{array}{cc}
   a & b \\
   c & d \\
 \end{array} \right)
 \left(\begin{array}{c}
   \omega_2 \\
   \omega_1 \\
 \end{array} \right),
 \;\;\;\;\;\;\;\;\;
 \left(\begin{array}{cc}
   a & b \\
   c & d \\
 \end{array} \right)
 \; \in \;
 \mathrm{SL}_2\left(\mathbb{Z}\right).
\eq
The transformation in eq.~(\ref{def_modular_transformation}) is called a modular transformation.
In terms of
\bq
 \tau' \; = \; \frac{\omega_2'}{\omega_1'},
 & &
 \tau \; = \; \frac{\omega_2}{\omega_1},
\eq
we have
\bq
\label{def_modular_transformation_tau}
 \tau' & = & \frac{a\tau+b}{c\tau+d}.
\eq
We denote the lattice generated by $(\tau,1)$ by $\Lambda_{\mathrm{norm}}$,
and the lattice generated by $(\tau',1)$ by $\Lambda_{\mathrm{norm}}'$.

Points in $\mathbb{C} / \Lambda$ can be identified with points on an elliptic curve.
Let $Z \in \mathbb{C} / \Lambda$. 
In going from a lattice generated by $(\omega_2,\omega_1)$ to a lattice generated by
$(\tau,1)$ we rescale all quantities by $1/\omega_1$.
Thus,
\bq
 z & = & \frac{Z}{\omega_1}
\eq
is the coordinate of our original point $Z \in \mathbb{C} / \Lambda$ in the normalised lattice $\mathbb{C} / \Lambda_{\mathrm{norm}}$.
Analogously,
\bq
 z' & = & \frac{Z}{\omega_1'}
\eq
is the coordinate of the point $Z$ in $\mathbb{C} / \Lambda_{\mathrm{norm}}'$.
The coordinates $z'$ and $z$ are related by $z'= (\omega_1/\omega_1') z$ or
\bq
\label{def_modular_transformation_z}
 z' & = & \frac{z}{c\tau +d}.
\eq
Eq.~(\ref{def_modular_transformation_tau}) and eq.~(\ref{def_modular_transformation_z}) give the transformation
of the modular parameter $\tau$ and of marked points $z$ on the elliptic curve under modular transformations, respectively.

For the differentials we have
\bq
 d\tau' \; = \; \frac{d\tau}{\left(c\tau+d\right)^2},
 & &
 dz' \; = \; \frac{dz}{c\tau +d} - \frac{c z d\tau}{\left(c\tau+d\right)^2}.
\eq

\subsection{Feynman integrals}

We consider a system of $r$ master integrals $J_1, \dots, J_r$, depending on $n$ kinematic variables $x_1, \dots, x_n$
within dimensional regularisation. The dimensional regularisation parameter is denoted by $\eps$.
We set $x=(x_1,\dots,x_n)$. We may think of $x$ as coordinates on a variety $B$, which we view as a base space.
We denote the vector of master integrals by $J=(J_1,\dots,J_r)^T$.
We may think of $J_1, \dots, J_r$ as a basis of a vector space $F$.
In mathematical terms we are considering a vector bundle with base space $B$ (of dimension $n$) 
and fibre $F$ (of dimension $r$).
The master integrals satisfy a differential equation with respect to the kinematic variables
\bq
 \left( d + A \right) J & = & 0.
\eq
$d$ denotes the exterior derivative on $B$ and 
$A$ is a $(r \times r)$-matrix, whose entries are differential one-forms.
In mathematical terms $A$ defines a (flat) connection on the fibre bundle.
The flatness of the connection follows from the integrability condition for $A$:
\bq
\label{integrability_condition}
 dA + A \wedge A & = & 0.
\eq
We say that the differential equation is in $\eps$-form \cite{Henn:2013pwa}, if the $(r \times r)$-matrix $A$ is of the form
\bq
\label{eps_form}
 A
 & = &
 \eps \sum\limits_{j=1}^{l} \; C_j \; \omega_j,
\eq
where
\begin{enumerate}
\item $C_j$ is a $(r \times r)$-matrix, whose entries are numbers $r_1+ir_2$ with $r_1,r_2 \in \mathbb{Q}$,
\item the only dependence on $\eps$ is given by the explicit prefactor,
\item the differential one-forms $\omega_j$ have only simple poles.
\end{enumerate}
We also write
\bq
 A & = &
 \sum\limits_{j=1}^n A_j \; dx_j.
\eq
If we change the basis of master integrals
\bq
\label{fibre_transformation}
 J' & = & U J,
\eq
where $U$ is a $(r \times r)$-matrix which may depend on $\eps$ and $x$,
the new connection matrix is given by
\bq
 A' & = & U A U^{-1} + U d U^{-1}.
\eq
If we perform a coordinate transformation on the base manifold
\bq
\label{base_transformation}
 x_i' & = & f_i\left(x\right), \;\;\;\;\;\;\;\;\; 1 \le i \le n,
\eq
the new connection matrix
\bq
 A' & = &
 \sum\limits_{j=1}^n A_j' \; dx_j'
\eq
is given by
\bq
 A_j'
 & = &
 \sum\limits_{i=1}^{n} A_i \; \frac{\partial x_i}{\partial x_j'}.
\eq
Under a combined transformation (a fibre transformation as in eq.~(\ref{fibre_transformation}) followed by
a coordinate transformation on the base manifold as in eq.~(\ref{base_transformation}))
we obtain
\bq
\label{fibre_base_transformation}
 A_j'
 & = &
 \sum\limits_{i=1}^{n} 
  \left( \frac{\partial x_i}{\partial x_j'} \right) 
  \left( U A_i U^{-1} + U \frac{\partial}{\partial x_i} U^{-1} \right).
\eq
If the differential equation is in $\eps$-form as in eq.~(\ref{eps_form}),
we may easily solve the differential equation in terms of iterated integrals \cite{Chen}.
Let
\bq
 \path & : & \left[a,b\right] \rightarrow B
\eq
be a path with start point ${x}_i=\path(a)$ and end point ${x}_f=\path(b)$.
Let us write
\bq
 f_j\left(\lambda\right) d\lambda & = & \path^\ast \omega_j
\eq
for the pull-backs to the interval $[a,b]$.
For $\lambda \in [a,b]$ the $d$-fold iterated integral 
of $\omega_1$, ..., $\omega_d$ along the path $\path$ is defined
by
\bq
\label{def_iterated_integral}
 I_{\path}\left(\omega_1,...,\omega_d;\lambda\right)
 & = &
 \int\limits_a^{\lambda} d\lambda_1 f_1\left(\lambda_1\right)
 \int\limits_a^{\lambda_1} d\lambda_2 f_2\left(\lambda_2\right)
 ...
 \int\limits_a^{\lambda_{d-1}} d\lambda_d f_d\left(\lambda_d\right).
\eq

\subsection{Elliptic Feynman integrals}
\label{subsect:elliptic_Feynman_integrals}

Let us now specialise to elliptic Feynman integrals.
We assume that the base space $B$ is obtained from a covering space 
of the moduli space $\mathcal{M}_{1,n'}$ of a curve of genus one with $n'$ marked points with $n' \ge n$ and where $(n'-n)$ points are held fixed.
As coordinates on the covering space of $\mathcal{M}_{1,n'}$ we may take
\bq
\label{standard_coordinates_full}
 \left(z_1,\dots,z_{n-1},z_n,\dots,z_{n'-1},\tau\right).
\eq
Translational invariance allows us to fix one marked point, which we take as $z_{n'}=0$.
Let us assume that we are only interested in the dependence on the coordinates
\bq
\label{standard_coordinates}
 \left(z_1,\dots,z_{n-1},\tau\right),
\eq
but not in the dependence on the coordinates $z_n,\dots,z_{n'-1}$. 
In order to distinguish them from $z_1,\dots,z_{n-1}$ we will write
\bq
 \beta_j & = & z_{j+n-1},
 \;\;\;\;\;\;\;\;\;
 1 \; \le \; j \; \le \; n'-n
\eq
and treat the $\beta_j$'s as additional parameters.
To make contact with our previous notation we have
\bq
 x_j & = & z_j, \;\;\;\;\;\;\;\;\;\;\;\; 1 \le j \le n-1,
 \nonumber \\
 x_n & = & \tau
\eq
and coordinates on the base space $B$ are given by eq.~(\ref{standard_coordinates}).
Choosing coordinates as in eq.~(\ref{standard_coordinates_full}) implies a choice
for the two periods $\omega_1$ and $\omega_2$ of the lattice $\Lambda$.

Let us further assume that for this choice of coordinates we have defined master integrals
$J=(J_1,\dots,J_r)^T$ such that the differential equation for the master integrals is
in $\eps$-form as in eq.~(\ref{eps_form}).
We may therefore solve the differential equation 
in terms of iterated integrals, say by integrating in the variable $\tau$.
For $|\qbar|$ small these iterated integrals have a rapidly converging series expansion in $\qbar$.
However, for
\bq
 \left| \qbar \right| & \lesssim & 1
\eq
the convergence is usually rather slow.
In these regions we would like to perform a modular transformation
\bq
\label{def_modular_transformation_M_1_n}
 \tau' & = & \frac{a\tau+b}{c\tau+d},
 \nonumber \\
 z_j' & = & \frac{z_j}{c\tau +d}, \;\;\;\;\;\;\;\;\;\;\;\; 1 \le j \le n-1,
 \nonumber \\
 \beta_j' & = & \frac{\beta_j}{c\tau +d}, \;\;\;\;\;\;\;\;\;\;\;\; 1 \le j \le n'-n,
\eq
such that $|\qbar'|$ is small.
Note that the additional parameters $\beta_j$ transform as well.


\section{The case $\mathcal{M}_{1,1}$}
\label{sect:M_1_1}

It is instructive to consider first the case where $B$ is a covering space of $\mathcal{M}_{1,1}$.
This case is simpler, as it does not yet have all complications.
The base space is one-dimensional and can be parametrised by a single coordinate $\tau$.

We consider the case, where all differential one-forms $\omega_j$ appearing in the 
differential equation~(\ref{eps_form}) are related to modular forms of some congruence subgroup $\Gamma$.
The definition of a congruence subgroup implies that there exists an $N$, such that
\bq
 \Gamma\left(N\right) & \subseteq & \Gamma.
\eq
This implies for the space of modular forms
\bq
 {\mathcal M}_k\left(\Gamma\right) & \subseteq & {\mathcal M}_k\left(\Gamma\left(N\right)\right).
\eq
It is therefore sufficient to restrict our attention to modular forms of 
the principal congruence subgroup $\Gamma(N)$.
For modular forms of level $N$ we set $\tau_N=\tau/N$ and 
\bq
 \qbar_N
 & = &
 e^{2\pi i \tau_N}
 \;\; = \;\;
 e^{\frac{2\pi i \tau}{N}}.
\eq
For a generic modular form $\eta$ of modular weight $k$ and level $N$ we set
\bq
\label{def_omega_modular}
 \omega^{\mathrm{modular}}\left(\eta\right)
 & = &
 2 \pi i \; \eta\left(\tau\right) \frac{d\tau}{N}
 \;\; = \;\;
 2 \pi i \; \eta\left(\tau\right) d\tau_N
 \;\; = \;\;
 \eta\left(\tau\right) \frac{d\qbar_N}{\qbar_N}.
\eq
If the modular form $\eta(\tau)$ has the $\qbar_N$-expansion
\bq
 \eta\left(\tau\right)
 & = &
 \sum\limits_{n=0}^\infty a_n \qbar_N^n,
\eq
we have
\bq
\label{differential_form_modular_form}
 \omega^{\mathrm{modular}}\left(\eta\right)
 & = &
 \sum\limits_{n=0}^\infty a_n \qbar_N^{n-1} d\qbar_N.
\eq
Let us now consider iterated integrals of modular forms.
It is customary to consider 
the integration path $\path$ from $\tau_i=i \infty$ to $\tau_f=\tau$.
For $\omega_1, \dots, \omega_d$ of the form as in eq.~(\ref{differential_form_modular_form}) 
and all of level $N$ we write
\bq
 \omega_j
 & = &
 \sum\limits_{n=0}^\infty a_{j,n} \qbar_N^{n-1} d\qbar_N.
\eq
For $a_{d,0}=0$ the iterated integral in eq.~(\ref{def_iterated_integral}) is given by
\bq
 I_{\path}\left(\omega_1,...,\omega_d;\tau\right)
 & = &
 \sum\limits_{i_1=1}^\infty \sum\limits_{i_2=1}^{i_1} \dots \sum\limits_{i_d=1}^{i_{d-1}}
  \qbar_N^{i_1}
  \frac{a_{1,i_1-i_2} \dots a_{d-1,i_{d-1}-i_d} a_{d,i_d}}
       {i_1 i_2 \cdot \dots \cdot i_d}.
\eq
If $|\qbar_N|$ is small this sum representation allows an efficient numerical evaluation of the iterated integral.
In order to arrive at the sum representation we repeatedly integrate
\bq
\label{basic_integration}
 \int\limits_0^{\qbar} \tilde{q}^{i-1} \; d\tilde{q}
 & = &
 \frac{1}{i} \qbar^i.
\eq
The condition $a_{d,0}=0$ is equivalent to the statement that $\omega_d$ 
vanishes on the cusp $\tau=i\infty$.
If $a_{d,0} \neq 0$ the iterated integral has a trailing zero.
With the help of the shuffle product an iterated integral with a trailing zero can be re-written
in terms of explicit prefactors $\ln(\qbar_N)$ and iterated integrals without trailing zeros \cite{Walden:2020odh}.

Let us now consider modular transformations.
It is convenient to introduce the $\slashoperator{\gamma}{k}$ operator acting on a modular form $\eta$ by
\bq
(\eta \slashoperator{\gamma}{k})(\tau) & = & (c\tau+d)^{-k} \cdot \eta(\gamma(\tau)).
\eq
For $\eta \in \mathcal{M}_k(\Gamma(N))$ we have
\begin{align}
\label{eta_modular_invariance}
 \eta \slashoperator{\gamma}{k} & = \eta,
 & 
 \gamma & \in \Gamma(N).
\end{align}
This is not yet too interesting, it only says that if $\eta$ is a modular form of weight $k$ for $\Gamma(N)$,
it transforms invariantly under the $\slashoperator{\gamma}{k}$ operation for any $\gamma \in \Gamma(N)$.
This statement is part of the definition of being a modular form of weight $k$ for $\Gamma(N)$.

We are interested in modular transformations from the full modular group $\mathrm{SL}_2({\mathbb Z})$,
and not just modular transformations restricted to the congruence subgroup $\Gamma(N)$.
Using the fact that $\Gamma(N)$ is a normal subgroup of $\mathrm{SL}_2({\mathbb Z})$ one can show that
we always have 
\begin{align}
\label{eta_modular_covariance}
 \eta \slashoperator{\gamma}{k} & \in \mathcal{M}_k(\Gamma(N)),
 & 
 \gamma & \in \mathrm{SL}_2(\mathbb{Z}),
\end{align}
e.g. we are not leaving the space of modular forms of weight $k$ for $\Gamma(N)$.
If $\eta$ is given as a polynomial in Eisenstein series for $\Gamma(N)$ we may 
compute $\eta \slashoperator{\gamma}{k}$ and express $\eta \slashoperator{\gamma}{k}$
again as a polynomial in Eisenstein series.
This is based on ref.~\cite{Broedel:2018iwv,Duhr:2019rrs} and reviewed in appendix~\ref{sect:Eisenstein}.

Let us now investigate the behaviour of iterated integrals of modular forms under modular transformations.
To see the problem it is sufficient to consider an iterated integral of depth one.
Let $\eta$ be a modular form of weight $k$ for $\Gamma(N)$ and define $\omega$ as in eq.~(\ref{def_omega_modular}).
For simplicity we assume that $\eta$ vanishes at the cusp $\tau=i\infty$.
We then have
\bq
 I_{\path}\left(\omega;\tau\right)
 & = &
 \frac{2\pi i}{N}
 \int\limits_{i \infty}^{\tau} \eta\left(\tilde{\tau}\right) d\tilde{\tau}
 \; = \;
 \sum\limits_{n=1}^\infty 
 \int\limits_{0}^{\qbar} a_n \tilde{q}_N^n \frac{d\tilde{q}_N}{\tilde{q}_N}
 \; = \;
 \sum\limits_{n=1}^\infty 
  \frac{a_{n}}{n}
  \qbar_N^{n}.
\eq
Let us now consider a coordinate transformation
\bq
 \tau & = & \gamma\left(\tau'\right) \; = \;
 \frac{a\tau'+b}{c\tau'+d},
 \;\;\;\;\;\;\;\;\;\;\;\;
 \gamma \; \in \; \mathrm{SL}_2({\mathbb Z}).
\eq
It is simpler to consider the inverse transformation here.
We have
\bq
 I_{\path}\left(\omega;\tau\right)
 & = &
 \frac{2\pi i}{N}
 \int\limits_{i \infty}^{\tau} \eta\left(\tilde{\tau}\right) d\tilde{\tau}
 \; = \;
 \frac{2\pi i}{N}
 \int\limits_{\gamma^{-1}\left(i \infty\right)}^{\gamma^{-1}\left(\tau\right)} 
  \eta\left(\gamma\left(\tilde{\tau}'\right)\right) \frac{d\tilde{\tau}'}{\left(c\tilde{\tau}'+d\right)^2}
 \nonumber \\
 & = &
 \frac{2\pi i}{N}
 \int\limits_{\gamma^{-1}\left(i \infty\right)}^{\gamma^{-1}\left(\tau\right)} 
 \left(c\tilde{\tau}'+d\right)^{k-2} 
 (\eta \slashoperator{\gamma}{k})(\tilde{\tau}')
 \; d\tilde{\tau}'.
\eq
$(\eta \slashoperator{\gamma}{k})(\tilde{\tau}')$ is again a modular form for $\Gamma(N)$, this is fine.
However, we picked up a factor $(c\tilde{\tau}'+d)^{k-2}$.
Only for the modular weight $k=2$ this factor is absent.
In general we leave the class of integrands constructed purely from modular forms.
For $k>2$ this can still be tolerated and will lead in the conversion of iterated integrals
to a sum representation to a generalisation of eq.~(\ref{basic_integration}) to integrals of the form
\bq
\label{log_integration}
 \int\limits_0^{\qbar} \tilde{q}^{i-1} \ln^j\left(\tilde{q}\right) d\tilde{q}.
\eq
However, for $k<2$ we obtain the automorphic factor $(c\tilde{\tau}'+d)$ in the denominator.
For this reason, the discussion in \cite{Manin:2005,Duhr:2019rrs} is restricted to modular weight $k \ge 2$
(and $0 \le j < k-1$ in eq.~(\ref{log_integration})).

We seek a better solution. In particular we would like to stay within the original class of functions.
For the case at hand we would like to stay within the class of iterated integrals of modular forms.
This can be achieved by a simultaneous transformation of the coordinate $\tau$ in the base variety
and a change of basis in the fibre.
This is best explained by an example.
The simplest example is the equal mass sunrise integral, consisting of three master integrals.
We start from
\bq
\label{def_sunrise}
 S_{\nu_1 \nu_2 \nu_3}\left( \eps, x \right)
 =
 \left(-1\right)^{\nu_{123}}
 e^{2\gamma_E \eps}
 \left(m^2\right)^{\nu_{123}-D}
 \int \frac{d^Dk_1}{i \pi^{\frac{D}{2}}} \frac{d^Dk_2}{i \pi^{\frac{D}{2}}}
 \frac{1}{D_1^{\nu_1} D_2^{\nu_2} D_3^{\nu_3}},
\eq
with the propagators
\bq
\label{def_propagators}
 D_1=k_1^2-m^2, \hspace{0.3cm}  
 D_2 = (k_1-k_2)^2-m^2, \hspace{0.3cm} 
 D_3 = (p-k_2)^2-m^2
\eq
and $D=2-2\eps$, $x=p^2/m^2$ and $\nu_{123}=\nu_1+\nu_2+\nu_3$.
$\gamma_E$ denotes Euler's constant.
From the maximal cut of the sunrise integral we obtain the elliptic curve
as a quartic polynomial $P(w,z)=0$:
\bq
\label{def_elliptic_curve}
 E
 & : &
 w^2 - z
       \left(z + 4 \right) 
       \left[z^2 + 2 \left(1+x\right) z + \left(1-x\right)^2 \right]
 \; = \; 0.
\eq
Let $\omega_1$ and $\omega_2$ be two periods of this elliptic curve with $\mathrm{Im}(\omega_2/\omega_1)>0$.
We set $\tau=\omega_2/\omega_1$.
We denote the Wronskian by
\bq
\label{def_Wronskian}
 W
 & = & 
 \omega_{1} \frac{d}{dx} \omega_{2} - \omega_{2} \frac{d}{dx} \omega_{1}.
\eq
Defining the master integrals
as 
\bq
\label{def_basis}
 J_1
 & = &
 4 \eps^2 \; S_{110}\left(\eps,x\right),
 \nonumber \\
 J_2
 & = &
 \eps^2 \frac{\pi}{\omega_1} \; S_{111}\left(\eps,x\right),
 \nonumber \\
 J_3
 & = &
 \frac{1}{\eps} \frac{\omega_1^2}{2 \pi i W} \frac{d}{dx} J_2 
 + \frac{\omega_1^2}{2 \pi i W} \frac{\left(3x^2-10x-9\right)}{2x\left(x-1\right)\left(x-9\right)} J_2,
\eq
and changing the variable on the base manifold from $x=p^2/m^2$ to $\tau$
puts the differential equation for $J=(J_1,J_2,J_3)^T$ in $\eps$-form \cite{Adams:2018yfj}
\bq
\label{example_sunrise_equal_dgl}
\left( d + A \right) J & = & 0
\eq
with
\bq
\label{example_sunrise_equal_A}
 A & = &
 2 \pi i \; \eps
 \left( \begin{array}{ccc}
 0 & 0 & 0 \\
 0 & \eta_2\left(\tau\right) & \eta_0\left(\tau\right) \\
 \eta_3\left(\tau\right) & \eta_4\left(\tau\right) & \eta_2\left(\tau\right) \\
 \end{array} \right)
 d\tau,
\eq
where $\eta_k(\tau)$ denotes a modular form of modular weight $k$.
The modular form $\eta_0(\tau)$ of weight zero is a constant, which we simply denote by $\eta_0$.
The entries $A_{2,2}$ and $A_{3,3}$ are identical.
For this particular example, the $\eta_k(\tau)$'s are modular forms of $\Gamma_1(6)$ 
and therefore also modular forms of
$\Gamma(6)$.
The specific expressions for the $\eta_k(\tau)$'s are not relevant for the discussion here.
Expressions for the $\eta_k(\tau)$'s in terms of Eisenstein series
are given in ref.~\cite{Honemann:2018mrb}.

Let us now consider for
\bq
 \gamma\left(\tau\right)
 & = & 
 \frac{a\tau+b}{c\tau+d},
 \;\;\;\;\;\;\;\;\;\;\;\;
 \gamma \; \in \; \mathrm{SL}_2({\mathbb Z})
\eq
the combined transformation
\bq
\label{combined_trafo_sunrise}
 J' 
 & = &
 \left( \begin{array}{ccc}
 1 & 0 & 0 \\
 0 & \left(c\tau+d\right)^{-1} & 0 \\
 0 & \frac{c}{2\pi i \eps \eta_0} & \left(c\tau+d\right) \\
 \end{array} \right) J,
 \nonumber \\
 \tau'
 & = &
 \frac{a\tau+b}{c\tau+d}.
\eq
Working out the transformed differential equation according to eq.~(\ref{fibre_base_transformation}) we obtain
\bq
\left( d + A' \right) J' & = & 0
\eq
with
\bq
 A' & = &
 2 \pi i \; \eps
 \left( \begin{array}{ccc}
 0 & 0 & 0 \\
 0 & (\eta_2 \slashoperator{\gamma^{-1}}{2})(\tau') & (\eta_0 \slashoperator{\gamma^{-1}}{0})(\tau') \\
 (\eta_3 \slashoperator{\gamma^{-1}}{3})(\tau') & (\eta_4 \slashoperator{\gamma^{-1}}{4})(\tau') & (\eta_2 \slashoperator{\gamma^{-1}}{2})(\tau') \\
 \end{array} \right)
 d\tau'.
\eq
We have
\bq
 \eta_k \slashoperator{\gamma^{-1}}{k}
 & \in & 
 \mathcal{M}_k(\Gamma(6))
\eq
and therefore we don't leave the space of modular forms with the combined transformation of eq.~(\ref{combined_trafo_sunrise}).
The transformed system may therefore again be solved 
for any $\gamma \in \mathrm{SL}_2({\mathbb Z})$
in terms of iterated integrals of modular forms.
In particular, we achieved that terms of the form as in eq.~(\ref{log_integration}) do not occur.
For this example and a few selected modular transformations this has been worked out in detail \cite{Honemann:2018mrb,Duhr:2019rrs}.

The fact that we need to redefine the master integrals is not too surprising. Let's look at $J_2$.
We originally defined $J_2$ by
\bq
 J_2
 & = &
 \eps^2 \frac{\pi}{\omega_1} \; S_{111}\left(\eps,x\right),
\eq
i.e. we rescaled $S_{111}$ (up to a constant) by $1/\omega_1$.
This definition is tied to our initial choice of periods.
Noting that the automorphic factor $(c\tau+d)$ is nothing than the ratio of two periods
\bq
 c \tau + d & = & 
 \frac{\omega_1'}{\omega_1},
\eq
we find that $J_2'$ is given by
\bq
 J_2'
 & = &
 \eps^2 \frac{\pi}{\omega_1'} \; S_{111}\left(\eps,x\right).
\eq


\section{The case $\mathcal{M}_{1,n'}$}
\label{sect:M_1_n}

We now consider the case where the base manifold is higher dimensional (i.e. the Feynman integrals depend
on more than one kinematic variable).
We take $B$ to be the space described in section~\ref{subsect:elliptic_Feynman_integrals}, parametrised by coordinates
\bq
 \left(z_1,\dots,z_{n-1},\tau\right).
\eq
We allow additional parameters $\beta_j$ (with $1 \le j \le n'-n$).
A modular transformation acts on these coordinates and the parameters $\beta_j$ as in eq.~(\ref{def_modular_transformation_M_1_n}).

We enlarge the set of differential one-forms $\omega$, 
which may appear in the differential equation for the Feynman integrals.
In addition to the differential one-forms related to modular forms as in eq.~(\ref{def_omega_modular})
we allow differential one-forms, which not only depend on $\tau$ but also on the other coordinates
$z_1, \dots, z_{n-1}$.
The simplest example is
\bq
\label{omega_Kronecker_simple}
 \omega_k\left(z_j,\tau\right)
 & = &
 \left(2\pi i\right)^{2-k}
 \left[
  g^{(k-1)}\left( z_j, \tau\right) d z_j + \left(k-1\right) g^{(k)}\left( z_j, \tau\right) \frac{d\tau}{2\pi i}
 \right].
\eq
The functions $g^{(k)}(z,\tau)$ are obtained from the expansion of the Kronecker function and reviewed in 
appendix~\ref{sect:Kronecker}.
We can be a little bit more general than eq.~(\ref{omega_Kronecker_simple}):
Let $K \in {\mathbb N}$ and $L(z)$ a linear function of $z_1, \dots, z_{n-1}$:
\bq
\label{linear_function}
 L\left(z\right)
 & = &
 \sum\limits_{j=1}^{n-1} \alpha_j z_j + \beta.
\eq
The generalisation of eq.~(\ref{omega_Kronecker_simple}) which we would like to consider is 
\bq
\label{omega_kronecker_general}
 \omega_k\left( L\left(z\right), K \tau\right)
 = 
 \left(2\pi i\right)^{2-k}
 \left[
  g^{(k-1)}\left( L\left(z\right), K \tau\right) d L\left(z\right) + K \left(k-1\right) g^{(k)}\left( L\left(z\right), K \tau\right) \frac{d\tau}{2\pi i}
 \right].
\eq
The differential one-form $\omega_k(L(z), K \tau)$ is closed
\bq
 d \omega_k\left(L\left(z\right), K \tau\right) & = & 0.
\eq
With the help of eq.~(\ref{relation_partial_derivatives}) it is not too difficult to prove this.

We may always reduce the case $K>1$ to the case $K=1$ with help of
\bq
\label{omega_reduce_K}
 \omega_k\left(L\left(z\right),K \tau\right)
 & = &
 \sum\limits_{l=0}^{K-1}
 \omega_k\left(\frac{L\left(z\right)+l}{K},\tau\right).
\eq
It is therefore sufficient to focus on the case $K=1$.

Let us now investigate the behaviour of $\omega_k(L(z), \tau)$ under a modular transformation.
The coordinates transform as in eq.~(\ref{def_modular_transformation_M_1_n}).
We assume that the parameter $\beta$ in eq.~(\ref{linear_function}) transforms as
\bq 
\label{trafo_constant}
 \beta'
 & = & 
 \frac{\beta}{c\tau+d}.
\eq
We may view $\beta$ as being a further marked point in a higher dimensional space
$\mathcal{M}_{1,n'}$ with $n'>n$.
With eq.~(\ref{trafo_constant}) we have
\bq
 L'\left(z'\right)
 & = &
 \sum\limits_{j=1}^{n-1} \alpha_j z_j' + \beta'
 \; = \;
 \frac{L\left(z\right)}{c\tau+d}.
\eq
We find
\bq
\label{trafo_omega_k}
 \omega_k\left(L'\left(z'\right),\tau'\right)
 & = &
 \left(c\tau +d \right)^{k-2}
 \sum\limits_{j=0}^k
 \frac{1}{j!}
 \left( \frac{c L\left(z\right)}{c\tau+d} \right)^j
 \omega_{k-j}\left(L\left(z\right),\tau\right).
\eq
The new feature are the additional terms with $j>0$.
They spoil the nice transformation properties under modular transformations.
Let us investigate how these terms disappear.

We might be tempted to try to absorb these terms into a redefinition of the master integrals.
However, this is not the way to proceed.
First of all, if we try to construct a suitable transformation matrix $U$ we will need in the
transformation matrix integrals of these terms. 
Secondly, a system of elliptic Feynman integral will usually contain 
in sub-sectors non-elliptic Feynman integrals as well.
These can be defined without any reference to a period of an elliptic curve
and we would not expect that a redefinition of these non-elliptic master integrals is necessary.

The mechanism how these terms cancel is different.
We illustrate it with a simple toy example.
Consider a system with one master integral ($r=1$), depending on two kinematic variables
$(z,\tau)$ and differential equation
\bq
\label{toy_example}
 \left( d+A \right) J & = & 0,
 \nonumber \\
 & & A \; = \; \eps \left[ \omega_2\left(z,\tau\right) - 2 \omega_2\left(z,2\tau\right) \right].
\eq
Using the periodicity
\bq
 \omega_{k}\left(L\left(z\right)+1,\tau\right)
 & = &
 \omega_{k}\left(L\left(z\right),\tau\right)
\eq
and eq.~(\ref{omega_reduce_K}) we may write
\bq
 \omega_2\left(z,\tau\right) - 2 \omega_2\left(z,2\tau\right)
 & = &
 \frac{1}{4} \omega_2\left(z-1,\tau\right)
 + \frac{1}{2} \omega_2\left(z,\tau\right)
 + \frac{1}{4} \omega_2\left(z+1,\tau\right)
 \nonumber \\
 & &
 - 2 \omega_2\left(\frac{z}{2},\tau\right)
 - \omega_2\left(\frac{z-1}{2},\tau\right)
 - \omega_2\left(\frac{z+1}{2},\tau\right).
\eq
On the right-hand side we may complete the lower weight terms, as the sum of all lower weight terms adds up to zero:
\bq
\lefteqn{
 \omega_2\left(z,\tau\right) - 2 \omega_2\left(z,2\tau\right)
 = } & &
 \\
 & &
 \frac{1}{4} \left[ \omega_2\left(z-1,\tau\right) 
                    + \frac{c\left(z-1\right)}{c\tau+d} \omega_1\left(z-1,\tau\right) 
                    + \frac{1}{2} \left( \frac{c\left(z-1\right)}{c\tau+d} \right)^2 \omega_0\left(z-1,\tau\right) 
             \right]
 \nonumber \\
 & &
 + \frac{1}{2} \left[ \omega_2\left(z,\tau\right)
                    + \frac{cz}{c\tau+d} \omega_1\left(z,\tau\right) 
                    + \frac{1}{2} \left( \frac{cz}{c\tau+d} \right)^2 \omega_0\left(z,\tau\right) 
               \right]
 \nonumber \\
 & &
 + \frac{1}{4} \left[ \omega_2\left(z+1,\tau\right)
                    + \frac{c\left(z+1\right)}{c\tau+d} \omega_1\left(z+1,\tau\right) 
                    + \frac{1}{2} \left( \frac{c\left(z+1\right)}{c\tau+d} \right)^2 \omega_0\left(z+1,\tau\right) 
               \right]
 \nonumber \\
 & &
 - 2 \left[ \omega_2\left(\frac{z}{2},\tau\right)
                    + \frac{cz}{2\left(c\tau+d\right)} \omega_1\left(\frac{z}{2},\tau\right) 
                    + \frac{1}{2} \left( \frac{cz}{2 \left(c\tau+d\right)} \right)^2 \omega_0\left(\frac{z}{2},\tau\right) 
     \right]
 \nonumber \\
 & &
 - \left[ \omega_2\left(\frac{z-1}{2},\tau\right)
                    + \frac{c\left(z-1\right)}{2\left(c\tau+d\right)} \omega_1\left(\frac{z-1}{2},\tau\right) 
                    + \frac{1}{2} \left( \frac{c\left(z-1\right)}{2 \left(c\tau+d\right)} \right)^2 \omega_0\left(\frac{z-1}{2},\tau\right) 
   \right]
 \nonumber \\
 & &
 - \left[ \omega_2\left(\frac{z+1}{2},\tau\right)
                    + \frac{c\left(z+1\right)}{2\left(c\tau+d\right)} \omega_1\left(\frac{z+1}{2},\tau\right) 
                    + \frac{1}{2} \left( \frac{c\left(z+1\right)}{2 \left(c\tau+d\right)} \right)^2 \omega_0\left(\frac{z+1}{2},\tau\right) 
   \right].
 \nonumber
\eq
Let us verify that the lower weight terms drop out: At weight zero we have
\bq
 \omega_0\left(L\left(z\right),\tau\right)
 & = & 
 - 2 \pi i \; d\tau
\eq
and
\bq
 \frac{1}{4} \left(z-1\right)^2
 + \frac{1}{2} z^2
 + \frac{1}{4} \left(z+1\right)^2
 -2 \left( \frac{z}{2} \right)^2
 - \left( \frac{z-1}{2} \right)^2
 - \left( \frac{z+1}{2} \right)^2
 & = &
 0.
\eq
At weight one we have
\bq
 \omega_1\left(L\left(z\right),\tau\right)
 & = & 
 2 \pi i \; dL\left(z\right)
\eq
and
\bq
 \frac{1}{4} \left(z-1\right) dz
 + \frac{1}{2} z dz
 + \frac{1}{4} \left(z+1\right) dz
 -2 \left( \frac{z}{2} \right) \frac{dz}{2}
 - \left( \frac{z-1}{2} \right) \frac{dz}{2}
 - \left( \frac{z+1}{2} \right) \frac{dz}{2}
 & = &
 0.
\eq
Please note that
\bq
 \omega_{k}\left(z+1,\tau\right)
 & = &
 \omega_{k}\left(z,\tau\right),
\eq
but in general
\bq
 \omega_{k}\left(\frac{z+1}{c\tau+d},\frac{a\tau+b}{c\tau+d}\right)
 & \neq &
 \omega_{k}\left(\frac{z}{c\tau+d},\frac{a\tau+b}{c\tau+d}\right).
\eq
For the system of eq.~(\ref{toy_example}) we obtain under a modular transformation $J'=J$ and
\bq
 \left( d+A' \right) J' & = & 0,
\eq
with
\bq
 A' & = & 
 \eps \left[ 
  \frac{1}{4} \omega_2\left(z'-\beta',\tau'\right)
  + \frac{1}{2} \omega_2\left(z',\tau'\right)
  + \frac{1}{4} \omega_2\left(z'+\beta',\tau'\right)
 \right. \nonumber \\
 & & \left.
  - 2 \omega_2\left(\frac{1}{2} z',\tau'\right)
  - \omega_2\left(\frac{1}{2} \left(z'-\beta'\right),\tau'\right)
  - \omega_2\left(\frac{1}{2} \left(z'+\beta'\right),\tau'\right)
 \right],
 \nonumber \\
 \beta' & = & 
 \frac{1}{c\tau+d}.
\eq
In the transformed equation no terms of lower weight appear.

Let us stress that we started in eq.~(\ref{toy_example}) from a system, which has nice modular transformation
properties.
This is not true of all systems.
Consider as a counter-example as before $r=1$ and $n=2$ with coordinates $(z,\tau)$, but now
\bq
\label{toy_example2}
 \left( d+A \right) J & = & 0,
 \nonumber \\
 & & A \; = \; \eps \omega_2\left(z,\tau\right).
\eq
This system does not transform nicely under modular transformations, as terms of weight zero and one
remain in the transformed differential equation.
Let us say that the system in eq.~(\ref{toy_example}) is modular, while the system in eq.~(\ref{toy_example2}) is not modular.

On physical grounds we expect that it should not matter which periods we choose for our elliptic curve.
If we find for a particular choice of periods $\omega_1$ and $\omega_2$ a differential equation, where each entry
of the connection matrix $A$ is a linear combination of terms as in 
eq.~(\ref{omega_kronecker_general}) or eq.~(\ref{def_omega_modular}),
we expect this to be the case for any other choice
of periods $\omega_1'$ and $\omega_2'$ as well.
This puts some constraints on the entries of $A$.
We call these constraints the modularity constraints.
The example in eq.~(\ref{toy_example}) satisfies the modularity constraints, while the example in eq.~(\ref{toy_example2}) does not.

Of course, the connection matrix $A$ has to satisfy the integrability condition of eq.~(\ref{integrability_condition}) as well.
The modularity constraints are additional constraints on top of this. 
This is easily seen from the examples in eq.~(\ref{toy_example}) and eq.~(\ref{toy_example2}):
Both examples satisfy the integrability condition trivially.

Let us now formalise this:
Let $\mathbb{F}$ be a field. Typically we take $\mathbb{F}$ to be $\mathbb{Q}$ or $\mathbb{Q}$ 
with some algebraic numbers adjoint.
We think of $\mathbb{F}$ as the fields of constants (of weight zero).
For the action of $\gamma$ on $L(z)$ we set
\bq
 \gamma\left(L\left(z\right)\right) 
 & = &
 \frac{L\left(z\right)}{c\tau+d}.
\eq
For a fixed modular weight $k$ consider the linear combination
\bq
\label{def_linear_combination}
 \omega\left(L_1(z),\dots,L_m(z),\tau\right)
 & = &
 \sum\limits_{j=1}^m C_j \; \omega_k\left(L_j\left(z\right),\tau\right),
 \;\;\;\;\;\;\;\;\;
 C_j \; \in \; {\mathbb F}.
\eq
An example for $\mathbb{F}=\mathbb{Q}[i,\sqrt{3}]$ would be
\bq
\label{def_linear_combination_example}
 \omega\left(2z_1,z_1+z_2,\tau\right)
 & = &
 \frac{i}{\sqrt{3}} \left[ \omega_4\left(2z_1,\tau\right) + 7 \omega_4\left(z_1+z_2,\tau\right) \right].
\eq
We define the action of the $\slashoperator{\gamma}{k}$ operator on $\omega$ as in eq.~(\ref{def_linear_combination})
by
\bq
(\omega \slashoperator{\gamma}{k})(L_1(z),\dots,L_m(z),\tau) 
 & = & 
 (c\tau+d)^{2-k} \cdot \omega(\gamma(L_1(z)),\dots,\gamma(L_m(z)),\gamma(\tau)).
\eq
The additional factor $(c\tau+d)^{2}$ comes from the fact that we are considering the transformation of a differential one-form, not a
function (see also eq.~(\ref{trafo_omega_k})).
We say that $\omega$ is modular invariant for $\Gamma(N)$ if
\begin{align}
\label{omega_modular_invariance}
 \omega \slashoperator{\gamma}{k} & = \omega,
 & 
 \gamma & \in \Gamma(N).
\end{align}
Let ${\mathcal M}_k^{\mathrm{elliptic}}$ be a $\mathbb{F}$-vector space generated by elements $\omega$ of the form 
as in eq.~(\ref{def_linear_combination}) and eq.~(\ref{def_omega_modular}) and of modular weight $k$.
For example, for $\omega(2z_1,z_1+z_2,\tau)$ defined in eq.~(\ref{def_linear_combination_example}) we have
\bq
 \omega\left(2z_1,z_1+z_2,\tau\right)
 & \in &
 {\mathcal M}_4^{\mathrm{elliptic}}.
\eq
Let $\omega \in {\mathcal M}_k^{\mathrm{elliptic}}$.
We say that $\omega$ is modular covariant with respect to ${\mathcal M}_k^{\mathrm{elliptic}}$ if
\begin{align}
\label{omega_modular_covariance}
 \omega \slashoperator{\gamma}{k} & \in {\mathcal M}_k^{\mathrm{elliptic}},
 & 
 \gamma & \in \mathrm{SL}_2({\mathbb Z}).
\end{align}
Eq.~(\ref{omega_modular_invariance}) and eq.~(\ref{omega_modular_covariance}) are the analogues
of eq.~(\ref{eta_modular_invariance}) and eq.~(\ref{eta_modular_covariance}) for modular forms.
Eq.~(\ref{omega_modular_covariance}) is weaker than eq.~(\ref{omega_modular_invariance}):
Eq.~(\ref{omega_modular_covariance}) states that we stay with the
$\slashoperator{\gamma}{k}$ operation inside the space ${\mathcal M}_k^{\mathrm{elliptic}}$, while
eq.~(\ref{eta_modular_invariance}) requires that $\omega$ is invariant under the 
$\slashoperator{\gamma}{k}$ operation.
Common to both definitions is the fact, 
that terms of lower modular weight $k'<k$ multiplied by coordinate dependent prefactors are absent.

Let us now consider a differential equation for a system of Feynman integrals.
We first define
\bq
 {\mathcal M}_{\bullet}^{\mathrm{elliptic}}
 & = & 
 \bigoplus\limits_{k=0}^{\infty} {\mathcal M}_{k}^{\mathrm{elliptic}}
\eq
and
\bq
 F_{k_{\mathrm{max}}} {\mathcal M}_{\bullet}^{\mathrm{elliptic}}
 & = & 
 \bigoplus\limits_{k=0}^{k_{\mathrm{max}}} {\mathcal M}_{k}^{\mathrm{elliptic}}.
\eq
${\mathcal M}_{\bullet}^{\mathrm{elliptic}}$ is the $\mathbb{F}$-vector space generated by elements $\omega$ 
of the form as in eq.~(\ref{def_linear_combination}) and eq.~(\ref{def_omega_modular}) and 
arbitrary modular weight $k$.
$F_{k_{\mathrm{max}}} {\mathcal M}_{\bullet}^{\mathrm{elliptic}}$ is the 
$\mathbb{F}$-vector space generated by elements $\omega$ 
of the form as in eq.~(\ref{def_linear_combination}) and eq.~(\ref{def_omega_modular}) with
modular weights ranging from $0$ to $k_{\mathrm{max}}$.
(The notation stems from the mathematical concept of a filtration.)
As an example we have
\bq
 \omega_2\left(z,\tau\right) + 2 \omega_1\left(z,\tau\right) + 6 \omega_0\left(z,\tau\right)
 & \in &
 F_{2} {\mathcal M}_{\bullet}^{\mathrm{elliptic}},
\eq
e.g. linear combinations of terms of different modular weight are allowed.
Not allowed are linear combinations with non-constant coefficients, e.g.
\bq
 \omega_2\left(z,\tau\right) 
 + \frac{z}{\tau+1} \omega_1\left(z,\tau\right) 
 & \notin &
 F_{2} {\mathcal M}_{\bullet}^{\mathrm{elliptic}}.
\eq
 Assume that the entries $A_{ij}$ of the matrix $A$ satisfy
\bq
 A_{ij}
 & \in &
 F_{k_{\mathrm{max}}} {\mathcal M}_{\bullet}^{\mathrm{elliptic}}.
\eq
This means that $A$ contains only terms of modular weight $0, 1, \dots, k_{\mathrm{max}}$ with constant coefficients.
We say that the differential equation for a system of Feynman integrals is modular, if 
for any $\gamma \in \mathrm{SL}_2({\mathbb Z})$ there exists a fibre transformation such that
the same condition holds for the transformed differential equation:
\bq
\label{condition_A}
 A_{ij}'
 & \in & 
 F_{k_{\mathrm{max}}} {\mathcal M}_{\bullet}^{\mathrm{elliptic}}.
\eq
This ensures that for any $\gamma \in \mathrm{SL}_2({\mathbb Z})$ the matrix $A$ has a power expansion in 
$\bar{q}'$. In particular this implies that terms of the form
\bq
 \left( c \tau' + d \right)^{j-2} & = & \left( \frac{c}{2\pi i} \ln \bar{q}' + d \right)^{j-2}
\eq
are absent.

Note that the condition in eq.~(\ref{condition_A}) is weaker than eq.~(\ref{omega_modular_covariance}).
Eq.~(\ref{condition_A}) allows linear combinations of terms with different modular weight, albeit with 
constant coefficients.
Coordinate dependent coefficients are not allowed.
Please note that we should not require a stronger condition.
Even if we start from a differential equation with a matrix $A$, where each entry 
is homogeneous in the modular weight, we would like to allow a simple redefinition
of the master integrals, where we replace one master integral 
by a sum of this master integral with a constant multiple of another master integral.
This transformation will in general lead to a matrix $A'$ with entries of mixed modular weight.
Allowing entries of mixed modular weight (with constant coefficients) does not spoil the property to express
the Feynman integrals as iterated integrals with integrands from $F_{k_{\mathrm{max}}} {\mathcal M}_{\bullet}^{\mathrm{elliptic}}$.

Let us illustrate this with an example: We have seen in section~\ref{sect:M_1_1}
that the differential equation of the equal mass sunrise integral in eq.~(\ref{example_sunrise_equal_dgl}) 
with $A$ given by eq.~(\ref{example_sunrise_equal_A}) 
\bq
 A & = &
 2 \pi i \; \eps
 \left( \begin{array}{ccc}
 0 & 0 & 0 \\
 0 & \eta_2\left(\tau\right) & \eta_0\left(\tau\right) \\
 \eta_3\left(\tau\right) & \eta_4\left(\tau\right) & \eta_2\left(\tau\right) \\
 \end{array} \right)
 d\tau
\eq
is modular.
Of course, the system will remain modular, if we perform a trivial change of master integrals according to
\bq
 \left(\begin{array}{c}
  \tilde{J}_1 \\
  \tilde{J}_2 \\
  \tilde{J}_3 \\
 \end{array} \right)
 & = &
 \left(\begin{array}{ccc}
  1 & 0 & 0 \\
  0 & 1 & 0 \\
  0 & 1 & 1 \\
 \end{array} \right)
 \left(\begin{array}{c}
  J_1 \\
  J_2 \\
  J_3 \\
 \end{array} \right).
\eq
In the basis $(\tilde{J}_1,\tilde{J}_2,\tilde{J}_3)^T$ the transformed matrix $\tilde{A}$
is given by
\bq
 \tilde{A} & = &
 2 \pi i \; \eps
 \left( \begin{array}{ccc}
 0 & 0 & 0 \\
 0 & \eta_2\left(\tau\right)-\eta_0\left(\tau\right) & \eta_0\left(\tau\right) \\
 \eta_3\left(\tau\right) & \eta_4\left(\tau\right)-\eta_0\left(\tau\right) & \eta_2\left(\tau\right)+\eta_0\left(\tau\right) \\
 \end{array} \right)
 d\tau.
\eq
We see that the entries of the matrix $\tilde{A}$ are not homogeneous in the modular weight.


\section{An example}
\label{sect:example}

Let us now consider a non-trivial example: We show the modularity of the two-loop
sunrise system with three unequal masses.
This system has seven master integrals 
\bq
 J & = & (J_1,J_2,J_3,J_4,J_5,J_6,J_7)^T
\eq
and depends on three kinematic variables
$(z_1,z_2,\tau)$.
Thus $r=7$ and $n=3$.

In ref.~\cite{Bogner:2019lfa} it was shown that the differential equation can be put into an $\eps$-form.
We closely follow the notation of ref.~\cite{Bogner:2019lfa} and take the definition of the variables
$(z_1,z_2,\tau)$ and the definition of the master integrals $J=(J_1,J_2,J_3,J_4,J_5,J_6,J_7)^T$
from there.
The differential equation reads
\bq
\label{diff_eq_sunrise_unequal}
 \left( d + A \right) J & = & 0,
\eq
with
\bq
 A & = &
 \eps
 \left( \begin{array}{ccccccc}
 a_{11} & 0 & 0 & 0 & 0 & 0 & 0 \\
 0 & a_{22} & 0 & 0 & 0 & 0 & 0 \\
 0 & 0 & a_{33} & 0 & 0 & 0 & 0 \\
 0 & 0 & 0 & a_{44} & a_{45} & a_{46} & a_{47} \\
 a_{51} & a_{52} & a_{53} & a_{54} & a_{55} & a_{56} & a_{57} \\
 a_{61} & a_{62} & a_{63} & a_{64} & a_{65} & a_{66} & a_{67} \\
 a_{71} & a_{72} & a_{73} & a_{74} & a_{75} & a_{76} & a_{77} \\
 \end{array} \right).
\eq
In order to present the entries of $A$ in a compact form we 
introduce\footnote{There are small differences in the notation used here and ref.~\cite{Bogner:2019lfa}:
In this paper the $\omega_k$'s are defined with a prefactor $(2\pi i)^{2-k}$ (see eq.~(\ref{omega_kronecker_general}), 
in ref.~\cite{Bogner:2019lfa} they are defined with a prefactor $(2\pi)^{2-k}$.
In this paper we set $z_3=-z_1-z_2$, in ref.~\cite{Bogner:2019lfa} $z_3$ was defined by $z_3=1-z_1-z_2$.
}
$z_3=-z_1-z_2$ and a constant $\beta=1$.
We define (for arbitrary $\beta$)
\bq
 \Omega_k\left(z,\beta,\tau\right)
 & = &
 \frac{1}{2} \omega_k\left(z,\tau\right)
 + \frac{1}{4} \omega_k\left(z-\beta,\tau\right)
 + \frac{1}{4} \omega_k\left(z+\beta,\tau\right)
 \nonumber \\
 & &
 - 2 \left(k-1\right) \left[
 \omega_k\left(\frac{z}{2},\tau\right)
 + \frac{1}{2} \omega_k\left(\frac{z-\beta}{2},\tau\right)
 + \frac{1}{2} \omega_k\left(\frac{z+\beta}{2},\tau\right)
 \right].
\eq
For $\beta=1$ we have
\bq
 \Omega_k\left(z,1,\tau\right)
 & = &
 \omega_k\left(z,\tau\right)
 - 2 \left(k-1\right) \omega_k\left(z,2\tau\right).
\eq
We will encounter $\Omega_2(z,\beta,\tau)$ and $\Omega_3(z,\beta,\tau)$.
Under a modular transformation we have for $\beta=1$ and $\beta'=1/(c\tau+d)$
\bq
 \Omega_2\left(z',\beta',\tau'\right)
 & = &
 \Omega_2\left(z,\beta,\tau\right),
 \nonumber \\
 \Omega_3\left(z',\beta',\tau'\right)
 & = &
 \left(c\tau +d \right)
 \left[
 \Omega_3\left(z,\beta,\tau\right)
 + \frac{c L\left(z\right)}{c\tau+d} \Omega_2\left(z,\beta,\tau\right)
 \right].
\eq
For the entries $a_{ij}$ we also need two differential forms $\eta_2(\tau)$ and $\eta_4(\tau)$, which depend on $\tau$, but not on the $z_i$'s.
These are defined by
\bq
 \eta_2\left(\tau\right)
 \; = \;
 \left[ e_2\left(\tau\right) - 2  e_2\left(2\tau\right) \right] \frac{d\tau}{2\pi i},
 & &
 \eta_4\left(\tau\right)
 \; = \;
 \frac{1}{\left(2\pi i\right)^2} e_4\left(\tau\right) \frac{d\tau}{2\pi i},
\eq
where $e_k(\tau)$ denotes the standard Eisenstein series.
The Eisenstein series $e_k(\tau)$ are defined in the appendix in eq.~(\ref{def_e_k}).
We have
\bq
 e_2\left(\tau\right) - 2  e_2\left(2\tau\right) \; \in \; {\mathcal M}_2(\Gamma_0(2)),
 & &
 e_4\left(\tau\right) \; \in \; {\mathcal M}_4(\mathrm{SL}_2({\mathbb Z})).
\eq
For the entries of $A$ we 
have the following relations
\begin{align}
 a_{45} & = \frac{1}{24} a_{57},
 &
 a_{46} & = \frac{1}{8} a_{67},
 & 
 a_{33} & = a_{11} + a_{22},
 \nonumber \\
 a_{53} & = a_{11} + a_{22} - a_{51} - a_{52},
 &
 a_{56} & = 3 a_{65},
 &
 a_{77} & = a_{44},
 \nonumber \\
 a_{61} & = 2 a_{11} - a_{51},
 &
 a_{62} & = -2 a_{11} + a_{51},
 &
 a_{63} & = a_{11}-a_{22},
 \nonumber \\
 a_{75} & = \frac{1}{24} a_{54},
 & 
 a_{76} & = \frac{1}{8} a_{64},
 & &
\end{align}
and the following symmetries
\begin{align}
 a_{22}\left(z_1,z_2,z_3\right) & = a_{11}\left(z_2,z_1,z_3\right),
 &
 a_{52}\left(z_1,z_2,z_3\right) & = a_{51}\left(z_2,z_1,z_3\right),
 \nonumber \\
 a_{72}\left(z_1,z_2,z_3\right) & = a_{71}\left(z_2,z_1,z_3\right),
 &
 a_{73}\left(z_1,z_2,z_3\right) & = a_{71}\left(z_1,z_3,z_2\right).
\end{align}
Thus we need to specify only a few entries. 
We group them by modular weight.
\\
\\
Modular weight $0$: As $\omega_0(z,\tau)=-2\pi i d\tau$ is independent of $z$, we simply write $\omega_0(\tau)$:
\bq
 a_{4,7}
 & = &
 \omega_0\left(\tau\right).
\eq
Modular weight $1$:
\bq
 a_{5,7}
 & = &
 6 i \left[ \omega_1\left(z_1,\tau\right) + \omega_1\left(z_2,\tau\right) \right], 
 \nonumber \\
 a_{6,7}
 & = &
 2 i \left[ \omega_1\left(z_1,\tau\right) - \omega_1\left(z_2,\tau\right) \right].
\eq
Note that $\omega_1(z,\tau)=2\pi i dz$ is independent of $\tau$.
\\
\\
Modular weight $2$:
\bq
 a_{1,1}
 & = &
 -2 \left[ \Omega_2\left(z_1,\beta,\tau\right) - \Omega_2\left(z_3,\beta,\tau\right) \right],
 \nonumber \\
 a_{4,4}
 & = &
 \omega_2\left(z_1,\tau\right) + \omega_2\left(z_2,\tau\right) + \omega_2\left(z_3,\tau\right) 
 - \Omega_2\left(z_1,\beta,\tau\right) - \Omega_2\left(z_2,\beta,\tau\right) + 3 \Omega_2\left(z_3,\beta,\tau\right) 
 \nonumber \\
 & &
 - 6 \eta_2\left(\tau\right),
 \nonumber \\
 a_{5,1}
 & = &
 - 2 \left[ \Omega_2\left(z_1,\beta,\tau\right) - \Omega_2\left(z_2,\beta,\tau\right) - 2 \Omega_2\left(z_3,\beta,\tau\right) \right],
 \nonumber \\
 a_{5,5}
 & = &
 - 3 \omega_2\left(z_3,\tau\right) 
 - \Omega_2\left(z_1,\beta,\tau\right) - \Omega_2\left(z_2,\beta,\tau\right) + 3 \Omega_2\left(z_3,\beta,\tau\right) 
 - 6 \eta_2\left(\tau\right),
 \nonumber \\
 a_{6,5}
 & = &
 - \omega_2\left(z_1,\tau\right) + \omega_2\left(z_2,\tau\right),
 \nonumber \\
 a_{6,6}
 & = &
 - 2 \omega_2\left(z_1,\tau\right) - 2 \omega_2\left(z_2,\tau\right) + \omega_2\left(z_3,\tau\right) 
 - \Omega_2\left(z_1,\beta,\tau\right) - \Omega_2\left(z_2,\beta,\tau\right) + 3 \Omega_2\left(z_3,\beta,\tau\right) 
 \nonumber \\
 & &
 - 6 \eta_2\left(\tau\right),
\eq
Modular weight $3$:
\bq
 a_{5,4}
 & = &
 12 i \left[ \omega_3\left(z_1,\tau\right) + \omega_3\left(z_2,\tau\right) - 2 \omega_3\left(z_3,\tau\right) \right],
 \nonumber \\
 a_{6,4}
 & = &
 12 i \left[ \omega_3\left(z_1,\tau\right) - \omega_3\left(z_2,\tau\right) \right],
 \nonumber \\
 a_{7,1}
 & = &
 i \left[ 
  \Omega_3\left(z_1,\beta,\tau\right) - \Omega_3\left(z_2,\beta,\tau\right) + \Omega_3\left(z_3,\beta,\tau\right) 
 \right].
\eq
Modular weight $4$:
\bq
 a_{7,4}
 & = &
 12 \left[ \omega_4\left(z_1,\tau\right) + \omega_4\left(z_2,\tau\right) + \omega_4\left(z_3,\tau\right) - 6 \eta_4\left(\tau\right) \right].
\eq
Let us now discuss the behaviour of the system under a modular transformation
\bq
 \gamma
 \; = \;
 \left(\begin{array}{cc}
   a & b \\
   c & d \\
 \end{array} \right)
 & \in &
 \mathrm{SL}_2({\mathbb Z}).
\eq
The coordinate transform as
\bq
\label{trafo_base_unequal_sunrise}
 z_1' \; = \; \frac{z_1}{c\tau+d},
 \;\;\;\;\;\;
 z_2' \; = \; \frac{z_2}{c\tau+d},
 \;\;\;\;\;\;
 \tau' \; = \; \frac{a\tau+b}{c\tau+d}.
\eq
The constant $\beta=1$ transforms as
\bq
\label{trafo_constants_unequal_sunrise}
 \beta' \; = \; \frac{\beta}{c\tau+d},
\eq
so in general we will have $\beta' \neq 1$.
We set again $z_3'=-z_1'-z_2'$.
We also need to redefine the master integrals.
We set
\bq
\label{trafo_fibre_unequal_sunrise}
 J' & = U J,
\eq
where $U$ is given by
\bq
\label{def_U_trafo_fibre_unequal_sunrise}
 U & = &
 \left( \begin{array}{ccccccc}
 1 & 0 & 0 & 0 & 0 & 0 & 0 \\
 0 & 1 & 0 & 0 & 0 & 0 & 0 \\
 0 & 0 & 1 & 0 & 0 & 0 & 0 \\
 0 & 0 & 0 & \frac{1}{c\tau+d} & 0 & 0 & 0 \\
 0 & 0 & 0 & \frac{6 i c \left(z_1+z_2\right)}{c\tau+d} & 1 & 0 & 0 \\
 0 & 0 & 0 & \frac{2 i c \left(z_1-z_2\right)}{c\tau+d} & 0 & 1 & 0 \\
 0 & 0 & 0 & -\frac{c}{2\pi i \eps} + \frac{c^2\left(z_1^2+z_1z_2+z_2^2\right)}{c\tau+d} & -\frac{i c \left(z_1+z_2\right)}{4} & -\frac{i c \left(z_1-z_2\right)}{4} & c\tau+d \\
 \end{array} \right).
\eq
The transformation matrix $U$ is not too difficult to construct, if one starts from the assumption that the first elliptic master integral (i.e. $J_4$) should
be rescaled as
\bq
 J_4' & = & 
 \frac{\omega_1}{\omega_1'} J_4
 \; = \;
 \frac{1}{c\tau+d} J_4.
\eq
Under this combined transformation 
the differential equation for the transformed system reads then
\bq
 \left( d + A' \right) J' & = & 0,
\eq
where $A'$ is obtained (with one exception) from $A$ by replacing all unprimed variables with primed variables.
For example $a_{7,1}'$ is given by
\bq
 a_{7,1}'
 & = &
 i \left[ 
  \Omega_3\left(z_1',\beta',\tau'\right) - \Omega_3\left(z_2',\beta',\tau'\right) + \Omega_3\left(z_3',\beta',\tau'\right) 
 \right].
\eq
The only exception is $\eta_2(\tau)$.
For $\gamma \in \Gamma_0(2)$ the differential one-form $\eta_2(\tau)$ transforms into
$\eta_2(\tau')$.
For a general $\gamma \in \mathrm{SL}_2({\mathbb Z})$ let us set $b_2(\tau)=e_2(\tau) - 2  e_2(2\tau)$.
Then $\eta_2(\tau)$ is replaced by
\bq
 (b_2 \slashoperator{\gamma^{-1}}{2})(\tau') \; \frac{d\tau'}{2\pi i}.
\eq
$(b_2 \slashoperator{\gamma^{-1}}{2})(\tau')$ is again a modular form (for $\Gamma(2)$),
but not necessarily identical to $b_2(\tau')$.

It remains to work out $(b_2 \slashoperator{\gamma^{-1}}{2})(\tau')$.
To this aim we first express $b_2(\tau)$ in terms of Eisenstein series for $\Gamma(2)$.
We find
\bq
 b_2\left(\tau\right)
 & = &
 4 \left(2\pi i\right)^2 h_{2,2,0,1}\left(\tau\right),
\eq
where the Eisenstein series $h_{k,N,r,s}(\tau)$ are defined in appendix~\ref{sect:Eisenstein}.
The transformation law for $b_2(\tau)$ follows then from the transformation law for $h_{k,N,r,s}(\tau)$
given in eq.~(\ref{trafo_Eisenstein_h_1}).
We obtain
\bq
 (b_2 \slashoperator{\gamma^{-1}}{2})(\tau')
 & = &
 4 \left(2\pi i\right)^2 
 h_{2,2,b \bmod 2,d \bmod 2}\left(\tau'\right),
 \;\;\;\;\;\;\;\;\;
 \gamma^{-1} \; = \;
 \left( \begin{array}{rr}
  d & -b \\
 -c & a \\
 \end{array}
 \right).
\eq
Let us summarise: 
The combined transformation of eq.~(\ref{trafo_base_unequal_sunrise}), eq.~(\ref{trafo_constants_unequal_sunrise}) 
and eq.~(\ref{trafo_fibre_unequal_sunrise}) transforms the differential equation to a new differential equation.
The entries of the new $A'$ are drawn from the same set of differential one-forms as the entries of the old $A$:
In both cases this set is given by 
differential one-forms defined by eq.~(\ref{omega_kronecker_general}) and related to the coefficients of the Kronecker function 
and
differential one-forms defined by eq.~(\ref{def_omega_modular}) and related to 
modular forms of $\Gamma(2)$.
The transformed system may therefore again be solved for any $\gamma \in \mathrm{SL}_2({\mathbb Z})$
in terms of iterated integrals with these letters.

As a final comment let us remark that the original differential equation in eq.~(\ref{diff_eq_sunrise_unequal})
has two particular properties:
(i) Each entry $A_{ij}$ of the matrix $A$ is homogeneous in the modular weight 
and (ii) in the equal-mass limit $m_1=m_2=m_3$ the master integrals $J_5$ and $J_6$ go to zero.
The transformation of eq.~(\ref{trafo_base_unequal_sunrise}), eq.~(\ref{trafo_constants_unequal_sunrise}) 
and eq.~(\ref{trafo_fibre_unequal_sunrise}) 
preserves property (i), but not property (ii).
In the equal-mass case $J_5'$ becomes proportional to $J_4'$.
This is related to the entry $U_{5,4}$ of the transformation matrix $U$.
We may enforce property (ii) by changing $U_{5,4}$ to
\bq
 U_{5,4}
 & = &
 \frac{6 i c \left(z_1+z_2-\frac{2}{3}\right)}{c\tau+d},
\eq
e.g. by adding the term $-2/3$ inside the bracket.
This amounts to adding a constant multiple of $J_4'$ to $J_5'$.
The transformed system is again such that the entries of the new $A'$ 
are drawn from the same set of differential one-forms as the entries of the old $A$.
However, the entries of the new connection matrix $A'$ contain now terms of mixed modular weight.
This is an example why we only require condition~(\ref{condition_A}) for being modular.


\section{Discussion}
\label{sect:discussion}

Let us summarise what we obtained so far:
We assume that we start from a differential equation in $\eps$-form, where the entries of the matrix $A$
are linear combinations of differential one-forms 
as in eq.~(\ref{def_omega_modular}) or eq.~(\ref{omega_kronecker_general}).
Let us further assume that all modular forms entering eq.~(\ref{def_omega_modular}) are given
as products of Eisenstein series.
Ref.~\cite{Duhr:2019rrs} gives us explicit formulae for the transformation of Eisenstein series
of the principal congruence subgroup $\Gamma(N)$ under arbitrary modular transformations $\gamma \in \mathrm{SL}_2({\mathbb Z})$, which are reviewed in appendix~\ref{sect:Eisenstein}.
Thus we know the modular transformation laws of all building blocks.
These are given by eq.~(\ref{trafo_Eisenstein_h_1}) and eq.~(\ref{trafo_omega_k}).

In order to show that the system is modular we have to find 
a matrix $U$, defining a basis transformation in the fibre, such that the transformed system
satisfies eq.~(\ref{condition_A}).
The fibre transformation adjusts automorphic factors $(c\tau+d)$
(this is already required for the simplest case discussed in section~\ref{sect:M_1_1})
and redistributes terms of lower weights such that they rearrange in combinations
as given by the right-hand side of eq.~(\ref{trafo_omega_k}).

If the system is modular, we may 
solve the transformed differential equation for the Feynman integrals in terms of iterated integrals 
with integrands drawn from the same class of integrands as the original system.

This is helpful for the numerical evaluation of the iterated integrals.
The original system can be evaluated efficiently whenever $|\bar{q}| \ll 1$.
We assume that the boundary constants are known for the original system.
Numerical routines for the evaluation of these iterated integrals in the region $|\bar{q}| \ll 1$
are implemented in GiNaC \cite{Bauer:2000cp} and described in \cite{Walden:2020odh}.
For $|\bar{q}|$ close to one we would like to switch to new coordinates such that $|\bar{q}'| \ll 1$.
If the system is modular, the transformed system can again be solved in terms of iterated integrals
from the same class of integrands.
However, we might not yet know the new boundary constants.
From the relation
\bq
 J' & = & U J
\eq
and evaluating both sides in an intermediate region where both $|\bar{q}|$ and $|\bar{q}|'$ are small,
we may extract numerically the new boundary constants.
This is similar to methods described in \cite{Hidding:2020ytt}.
Evaluating all expressions to high precision, 
it is often possible with the help of the PSLQ algorithm \cite{Ferguson:1992}
to convert the numerically known new boundary constants to analytic expressions
as linear combinations of transcendental constants.
With the new boundary constants at hand, we obtain efficient evaluations of the Feynman integrals
in the new region where $|\bar{q}| \lesssim 1$, but $|\bar{q}'| \ll 1$.

The modular properties discussed in this paper can also be used for finding the original differential
equation in $\eps$-form.
A strategy for finding this differential equation may use an ansatz of appropriate terms 
and fixing the unknown coefficients by comparing $\bar{q}$-expansions.
Assuming that the differential equation should be modular, we may reduce
the number of terms in the ansatz by only allowing terms which lead to a modular differential equation.

Finally let us remark that the coordinate dependent basis transformation for the master integrals
discussed in eq.~(\ref{combined_trafo_sunrise}) and eq.~(\ref{def_U_trafo_fibre_unequal_sunrise})
provide non-trivial examples, where the transformed differential equation is again in $\eps$-form.
This also shows that there is no canonical choice for the master integrals.
Any definition of master integrals which puts the differential equation for a system of
elliptic Feynman integrals into an $\eps$-form will be based on a choice of a pair 
of periods for the elliptic curve.


\section{Conclusions}
\label{sect:conclusions}

In this paper we investigated the behaviour of elliptic Feynman integrals under modular transformations.
Modular transformations can be used to ensure that the nome squared is a small quantity.
More concretely, by a suitable modular transformation from the full modular group $\mathrm{SL}_2({\mathbb Z})$
we always can achieve $| \qbar | \le 0.0043$.
If the nome squared is small, we may evaluate the elliptic Feynman integrals efficiently through a $\bar{q}$-expansion.
Routines to do so are available within GiNaC \cite{Bauer:2000cp,Walden:2020odh}.

In this paper we investigated the question, whether we stay by a modular transformation within the same class
of iterated integrals.
The answer is yes, but only if we simultaneously transform the basis of master integrals as well.

This is different from ``ordinary'' Feynman integrals, which evaluate to multiple polylogarithms.
In the case of multiple polylogarithms we may use transformations of the arguments like $x'=1/x$
to transform a multiple polylogarithm into functions of the same class, 
which all have fast convergent series expansions \cite{Vollinga:2004sn}.
This can be done without redefining the master integrals.

This has practical implications: 
For multiple polylogarithms we may provide numerical evaluation routines for all values of the arguments.
If some arguments are outside the region of convergence of the sum representation, 
the numerical evaluation routines internally transform these arguments to the region of convergence 
and stay always within the class of multiple polylogarithms.

This is not the case for elliptic Feynman integrals: We may express elliptic Feynman integrals
in terms of iterated integrals with integrands given by
eq.~(\ref{def_omega_modular}) and eq.~(\ref{omega_kronecker_general}).
For $| \qbar |$ small, these iterated integrals are evaluated efficiently through a $\bar{q}$-expansion.
However, if we just consider this class of iterated integrals and their behaviour under modular transformations,
we find that we do not stay within the original class of iterated integrals.
Thus it is not possible to provide a ``black-box''-algorithm, which provides numerical evaluations for all
possible values of the arguments and always stays within this class of iterated integrals.
This does not exclude the possibility, that there is a larger class of iterated integrals where this is possible.
However, enlarging the class of iterated integrals would not be natural.
The natural solution is to consider a combined transformation, consisting of a modular transformation
of the variables of the base manifold and a basis transformation of the master integrals in the fibre.
Under such a combined transformation we stay within the original class of iterated integrals
with integrands of the form as in 
eq.~(\ref{def_omega_modular}) or eq.~(\ref{omega_kronecker_general}).

This is the main result of this paper: 
For elliptic Feynman integrals we should always consider
a modular transformation of the variables of the base manifold together with a redefinition of the master integrals.


\begin{appendix}


\section{Eisenstein series for $\Gamma(N)$}
\label{sect:Eisenstein}

Let $r, s$ be integers with $0 \le r,s <N$.
Following \cite{Broedel:2018iwv,Duhr:2019rrs} we define Eisenstein series $h_{k,N,r,s}(\tau$) for $\Gamma(N)$
by
\bq
\label{def_Eisenstein_h}
 h_{k,N,r,s}\left(\tau\right)
 & = &
 \sum\limits_{n=1}^\infty a_n \qbar^n_N.
\eq
For $n \ge 1$ the coefficients are given by
\bq
\label{coeff_h_a_n}
 a_n
 & = &
 \frac{1}{2 N^k}
 \sum\limits_{d|n}
 \sum\limits_{c_1=0}^{N-1}
 d^{k-1}
 \left[
   e^{\frac{2\pi i}{N} \left(r \frac{n}{d} - \left(s-d\right) c_1 \right)}
   + \left(-1\right)^k
   e^{-\frac{2\pi i}{N} \left(r \frac{n}{d} - \left(s+d\right) c_1 \right)}
 \right].
\eq
The constant term is given for $k\ge2$ by
\bq
 a_0
 & = & 
 - \frac{1}{2k} B_k\left(\frac{s}{N}\right),
\eq
where $B_k(x)$ is the $k$'th Bernoulli polynomial defined by
\bq
 \frac{t e^{x t}}{e^t-1}
 & = &
 \sum\limits_{k=0}^\infty \frac{B_k\left(x\right)}{k!} t^k.
\eq
For $k=1$ the constant term is given by
\bq
 a_0
 & = &
 \left\{
 \begin{array}{ll}
  \frac{1}{4} - \frac{s}{2N}, & s \neq 0, \\
 0, & (r,s) = (0,0), \\
 \frac{i}{4} \cot\left(\frac{r}{N} \pi \right),
 & \mbox{otherwise}.
 \end{array}
 \right.
\eq
With the exception of $(k,r,s) \neq (2,0,0)$ the $h_{k,N,r,s}(\tau)$ are Eisenstein series for $\Gamma(N)$:
\bq
 h_{k,N,r,s}\left(\tau\right) 
 & \in & 
 \mathcal{E}_k\left(\Gamma\left(N\right)\right).
\eq
For $(k,N,r,s)=(2,1,0,0)$ we have
\bq
 h_{2,1,0,0}\left(\tau\right)
 & = &
 \frac{1}{2\left(2\pi i\right)^2} e_2\left(\tau\right),
\eq
which is not a modular form.
$e_k(\tau)$ denotes the Eisenstein series
\bq
\label{def_e_k}
 e_k\left(\tau\right)
 & = &
 \sideset{}{_e}\sum\limits_{(n_1,n_2) \in {\mathbb Z}^2\backslash (0,0)} \frac{1}{\left(n_1 + n_2\tau \right)^k},
\eq
with the standard Eisenstein summation prescription understood (this is indicated by the sub-script $e$).

The Eisenstein series $h_{k,N,r,s}(\tau)$
transform under modular transformations 
\bq
 \gamma \; = \;
 \left(\begin{array}{cc} a & b \\ c & d \\ \end{array} \right)
 & \in &
 \mathrm{SL}_2\left({\mathbb Z}\right)
\eq
of the full modular group $\mathrm{SL}_2({\mathbb Z})$ as
\bq
\label{trafo_Eisenstein_h_1}
 h_{k,N,r,s}\left(\frac{a\tau+b}{c\tau+d}\right)
 & = &
 \left(c\tau +d\right)^k
 h_{k,N,rd+sb \bmod N,rc+sa \bmod N}\left(\tau\right),
\eq
or equivalently with the help of the $\slashoperator{\gamma}{k}$ operator
\bq
\label{trafo_Eisenstein_h_2}
 \left(h_{k,N,r,s}\slashoperator{\gamma}{k}\right)\left(\tau\right)
 & = &
 h_{k,N,rd+sb \bmod N,rc+sa \bmod N}\left(\tau\right).
\eq


\section{The Kronecker function}
\label{sect:Kronecker}

Let us define the theta function $\theta_1(z,\qbar)$ by
\bq
\theta_1\left(z,\qbar\right) 
 & = &
 -i \sum\limits_{n=-\infty}^\infty \left(-1\right)^n \qbar^{\frac{1}{2}\left(n+\frac{1}{2}\right)^2} e^{i \pi \left(2n+1\right)z},
\eq
and the Kronecker function $F(x,y,\tau)$ by
\bq
 F\left(x,y,\tau\right)
 & = &
 \theta_1'\left(0,\qbar\right) \frac{\theta_1\left(x+y, \qbar\right)}{\theta_1\left(x, \qbar\right)\theta_1\left(y, \qbar\right)}.
\eq
$\theta_1'$ denotes the derivative with respect to the first argument.
It is obvious from the definition that the Kronecker function is symmetric in $x$ and $y$.
We are interested in the Laurent expansion in one of these variables.
We define functions
$g^{(k)}(z,\tau)$ through
\bq
\label{def_g_n}
 F\left(z,\alpha,\tau\right)
 & = &
 \sum\limits_{k=0}^\infty g^{(k)}\left(z,\tau\right) \alpha^{k-1}.
\eq
The functions $g^{(k)}(z,\tau)$ will enter the definition of elliptic multiple polylogarithms.
Let us recall some of their properties \cite{Zagier:1991,Brown:2011,Broedel:2018qkq}.
When viewed as a function of $z$, the function $g^{(k)}(z,\tau)$ has only simple poles.
More concretely, the function $g^{(1)}(z,\tau)$ has a simple pole with unit residue at every point of the lattice.
For $k>1$ the function $g^{(k)}(z,\tau)$ has a simple pole only at those lattice points 
that do not lie on the real axis.
The (quasi-) periodicity properties are
\bq
 g^{(k)}\left(z+1,\tau\right) & = &  g^{(k)}\left(z,\tau\right),
 \nonumber \\
 g^{(k)}\left(z+\tau,\tau\right) & = &  
 \sum\limits_{j=0}^k \frac{\left(-2\pi i\right)^j}{j!} g^{(k-j)}\left(z,\tau\right).
\eq
We see that $g^{(k)}(z,\tau)$ is invariant under translations by $1$, but not by $\tau$.
The functions $g^{(k)}(z,\tau)$ have the symmetry
\bq
 g^{(k)}(-z,\tau)
 & = &
 \left(-1\right)^k g^{(k)}(z,\tau).
\eq
Let us introduce
\bq
 \mathrm{ELi}_{n;m}\left(\bar{u};\bar{v};\qbar\right) & = & 
 \sum\limits_{j=1}^\infty \sum\limits_{k=1}^\infty \; \frac{\bar{u}^j}{j^n} \frac{\bar{v}^k}{k^m} \qbar^{j k}
\eq
and the linear combinations
\bq
 \overline{\mathrm{E}}_{n;m}\left(\bar{u};\bar{v};\qbar\right) 
 & = &
  \mathrm{ELi}_{n;m}\left(\bar{u};\bar{v};\qbar\right)
  - \left(-1\right)^{n+m} \mathrm{ELi}_{n;m}\left(\bar{u}^{-1};\bar{v}^{-1};\qbar\right).
\eq
The functions $\overline{\mathrm{E}}_{n;m}$ are helpful for the $\qbar$-expansion of the functions $g^{(k)}(z,\tau)$.
Explicitly one has with $\qbar=\exp(2\pi i\tau)$ and $\bar{w}=\exp(2\pi i z)$
\bq
\label{g_n_explicit}
 g^{(0)}\left(z,\tau\right)
 & = & 1,
 \nonumber \\
 g^{(1)}\left(z,\tau\right)
 & = &
 - 2 \pi i \left[
                  \frac{1+\bar{w}}{2 \left(1-\bar{w}\right)}
                  + \overline{\mathrm{E}}_{0,0}\left(\bar{w};1;\qbar\right)
 \right],
 \nonumber \\
 g^{(k)}\left(z,\tau\right)
 & = &
 - \frac{\left(2\pi i\right)^k}{\left(k-1\right)!} 
 \left[
 - \frac{B_k}{k}
       + \overline{\mathrm{E}}_{0,1-k}\left(\bar{w};1;\qbar\right)
 \right],
 \;\;\;\;\;\;\;\;\;\;\;\;\;\;\;\;\;\;\;\;\;\;\;\;\;\;\;
 k > 1,
\eq
where $B_k$ denote the $k$-th Bernoulli number, defined by
\bq
 \frac{x}{e^x-1}
 & = &
 \sum\limits_{k=0}^\infty \frac{B_k}{k!} x^k.
\eq
It will be convenient to set $g^{(-1)}(z,\tau)=0$.
The partial derivatives satisfy
\bq
\label{relation_partial_derivatives}
 2 \pi i \frac{\partial}{\partial \tau} g^{(k-1)}\left(z,\tau\right)
 & = &
 \left(k-1\right) \frac{\partial}{\partial z} g^{(k)}\left(z,\tau\right).
\eq
A useful formula, which relates functions with argument $K \tau$ (with $K \in {\mathbb N}$) to functions with
argument $\tau$ reads
\bq
 g^{(k)}\left(z, K \tau\right)
 & = &
 \frac{1}{K} \sum\limits_{l=0}^{K-1} 
  g^{(k)}\left(\frac{z+l}{K},\tau\right).
\eq
Under modular transformations the functions $g^{(k)}(z,\tau)$ transform as
\bq
 g^{(k)}\left(\frac{z}{c\tau+d},\frac{a\tau+b}{c\tau+d}\right)
 & = &
 \left(c\tau +d \right)^k
 \sum\limits_{j=0}^k
 \frac{\left(2\pi i\right)^j}{j!}
 \left( \frac{c z}{c\tau+d} \right)^j
 g^{(k-j)}\left(z,\tau\right).
\eq

\end{appendix}

{\footnotesize
\bibliography{/home/stefanw/notes/biblio}
\bibliographystyle{/home/stefanw/latex-style/h-physrev5}
}

\end{document}